# Lattice Stellar Dynamics

*D. Syer*[1] *and S. Tremaine*[1,2]

[1]Canadian Institute for Theoretical Astrophysics, McLennan Labs, University of Toronto, 60 St. George Street, Toronto M5S 1A7, Canada

[2]Institute of Astronomy, Madingley Road, Cambridge CB3 0HA, England

# Summary

We describe a technique for solving the combined collisionless Boltzmann and Poisson equations in a discretised, or lattice, phase space. The time and the positions and velocities of 'particles' take on integer values, and the forces are rounded to the nearest integer. The equations of motion are symplectic. In the limit of high resolution, the lattice equations become the usual integro-differential equations of stellar dynamics. The technique complements other tools for solving those equations approximately, such as $N$-body simulation, or techniques based on phase-space grids. Equilibria are found in a variety of shapes and sizes. They are true equilibria in the sense that they do not evolve with time, even slowly, unlike existing $N$-body approximations to stellar systems, which are subject to two-body relaxation. They can also be 'tailor-made' in the sense that the mass distribution is constrained to be close to some pre-specified function. Their principal limitation is the amount of memory required to store the lattice, which in practice restricts the technique to modeling systems with a high degree of symmetry. We also develop a method for analysing the linear stability of collisionless systems, based on lattice equilibria as an unperturbed model.

# 1 Introduction

The structure of a collisionless stellar system is determined by the distribution function $f(\mathbf{r}, \mathbf{v}, t)$, which gives the density in phase space as a function of position $\mathbf{r}$, velocity $\mathbf{v}$ and time $t$. The spatial density of the system can be written

$$\rho(\mathbf{r}, t) = \int f(\mathbf{r}, \mathbf{v}, t) \, d^3\mathbf{v}. \tag{1}$$

The gravitational potential, $\Phi(\mathbf{r}, t)$, is determined by Poisson's equation:

$$\nabla^2 \Phi = 4\pi G \rho; \tag{2}$$

and the evolution of $f$ is determined by the collisionless Boltzmann equation

$$\frac{\partial f}{\partial t} + \mathbf{v} \cdot \boldsymbol{\nabla} f - \boldsymbol{\nabla}\Phi \cdot \frac{\partial f}{\partial \mathbf{v}} = 0, \qquad (3)$$

which is just a Lagrangian expression of conservation of phase-space density.

There exist exact solutions to equations (1–3) only in certain special cases (see for example Binney and Tremaine 1987). Approximate methods of solution include $N$-body simulations, phase-space methods, Schrödinger methods, and orbit-based techniques. The $N$-body approach is popular, in part because new algorithms such as tree codes, grid methods, and potential expansions have led to dramatic increases in efficiency (Sellwood 1987). Despite these improvements the number of particles, $N$, remains several orders of magnitude smaller than the number in real galactic-scale stellar systems. $N$-body simulations relax owing to stellar encounters on a timescale that is roughly proportional to $N$; hence the effects of encounters are much stronger in the simulation than in a real stellar system, and systems that are precisely in a steady state cannot be constructed.

The phase-space approach is essentially approximate 6-dimensional hydrodynamics, and suffers from technical difficulties (e.g. Goldstein, Cuperman and Lecar 1969, Inagaki, Nishida and Sellwood 1984, White 1986, Rasio, Shapiro and Teukolsky 1989) partly stemming from the tendency of the solutions of (3) to foliate on ever-decreasing scales—the phase fluid, $f$, evolves 'fingers' and becomes extremely complicated in a fine-grained sense within a few crossing times. We are uninterested in the fine-grained behaviour of $f$, and we would be satisfied to know what the system looked like on a scale small compared with its overall size, but large compared to the fingers. A second, more serious problem is that the amount of information that must be carried, even in a smooth phase fluid, is enormous in a 6-dimensional phase space, thus restricting simulations to systems having a reduced phase space owing to symmetry.

Schrödinger methods solve the Schrödinger equation with an artificially large value of Planck's constant (Widrow and Kaiser 1993). In this method the entire evolution of the stellar system is specified by the wavefunction, which depends on the 3 spatial coordinates and the time only (in contrast to the distribution function, which depends on 6 phase-space dimensions and time). However, the spatial resolution required for the wavefunction is much higher than for the distribution function, so the amount of information that must be carried is similar.

A final approach is to use orbit-based techniques, such as the linear-programming method of Schwarzschild (1979). A library of orbits and their contribution to the density field is compiled in some specified potential, and linear programming, or some other optimization algorithm such as maximum entropy, is used to assign weights to the orbits. This method is often slow to converge, and the answer can depend on the choice of the library of orbits. It is difficult to program, and is generally restricted to simulating equilibrium models. Schwarzschild (1979) used the method to provide an example of a triaxial equilibrium stellar system. It has been used with success by others, most notably for systems with high symmetry or self-similar systems (e.g. Kormendy and Richstone 1992, Schwarzschild 1993). A closely related technique employs phase-space basis functions rather than orbits as building blocks (e.g. Dejonghe 1989).

In this paper we investigate another approach to modelling stellar systems, based on a lattice phase space. There exist lattice-based calculations which take a similar approach to modeling the flow of gases (see Chen 1993 for a review). These methods can model a gas using only a few degrees of freedom in the velocity dimension, essentially because fluid dynamics depend mainly on the first and second velocity moments of $f$; but for stellar systems higher resolution in velocity space is required. Miller and Prendergast (1968) describe some stellar-dynamical simulations in a discrete phase space; essentially these were an approximation to $N$-body simulations. They were only able to run their simulations for a small number of crossing times. Although our approach is similar to Miller and Prendergast's, with the advantage of modern computers, we can now perform more interesting simulations.

The equations of lattice stellar dynamics are exact in a precise sense defined below and by Earn and Tremaine (1992). A particular attraction of lattice stellar dynamics is that we can construct truly time-independent solutions. Thus we shall focus on the problem of constructing self-consistent equilibrium systems.

## 2 Lattice Phase Space

In a lattice phase space, the motion of particles is restricted to a set of discrete points. If the lattice is rectangular, these points are separated by a constant distance $\Delta x$ in position, and $\Delta v$ in velocity. The minimum timescale on which a particle can change its position and stay on the lattice is $\Delta t = \Delta x/\Delta v$. Without loss of generality we may consider the positions and velocities of such particles to be specified by integer values, so that

$$\Delta x = \Delta v = \Delta t = 1. \qquad (4)$$

The mass of a particle is specified by a real number $m$. In order that the motion of a particle is restricted to the lattice, the acceleration that it feels must also be restricted to integer values.

We use a first-order modified Euler procedure to evolve particles in lattice phase space (higher order procedures are easy to derive and use but truncation error is not the limiting factor in our models). A particle at $(\mathbf{r}, \mathbf{v})$ moves to $(\mathbf{r}', \mathbf{v}')$ in one timestep according to:

$$\begin{aligned} \mathbf{v} \to \mathbf{v}' &= \mathbf{v} + \mathbf{a}, \\ \mathbf{r} \to \mathbf{r}' &= \mathbf{r} + \mathbf{v}', \end{aligned} \qquad (5)$$

where $\mathbf{a}$ is the (integerised) acceleration at the point $\mathbf{r}$. This procedure can be thought of as an integration algorithm with zero round-off error, and truncation error controlled by the resolution of the lattice (Earn and Tremaine 1992). The integration is exactly symplectic, in the sense defined by Earn and Tremaine (1992). The limit of a continuous system is approached as the resolution is increased; this corresponds to the limit of a large range in **r** and **v** with the convention (4). In this limit there is a conserved quantity corresponding to the energy of a particle, $E = \frac{1}{2}v^2 + \Phi(\mathbf{r})$, if the acceleration **a** can be derived from a potential $\Phi(\mathbf{r})$ (however the convergence is not smooth: as the resolution is increased the potential and force approach their continuous limits, but derivatives of the force do not).

To simulate a solution of the collisionless Boltzmann equation (3), we evolve a set of particles with masses, $m$, equal to the initial values of $f$ evaluated at the lattice sites. The equivalent of (3) is simply

$$m(\mathbf{r}', \mathbf{v}', t+1) = m(\mathbf{r}, \mathbf{v}, t), \tag{6}$$

with **a** given by

$$\mathbf{a} = [\,\boldsymbol{\nabla}\Phi(\mathbf{r})\,], \tag{7}$$

where [ ] is used to denote rounding to the nearest integer. The real forces, $\boldsymbol{\nabla}\Phi(\mathbf{r})$, are derived from the solution of (1) and (2) with $f$ replaced by the appropriate sum of point masses at the lattice sites. To minimise problems with rounding, the forces are calculated with greater precision (8-byte) than the positions and velocities (2- or 4-byte integers). The forces are calculated exactly at the grid positions using a fast Fourier transform, by doubling the dimensions of the grid on which the density is measured and filling the remainder with zeros (Press *et al.* 1992). This zero-padding requires extra storage and extra computing time (roughly doubling both for each space dimension). The extra storage is only in the space dimensions, however, and therefore is small compared to the storage required for the phase-space lattice; thus we have not implemented more sophisticated algorithms that require less storage (Sellwood 1987). So far the algorithm we have described is closely similar to the 'game' played by Miller and Prendergast (1968). Unlike those authors, we do not use a fixed grid in velocity space; we economise by storing the velocity information only where $m$ is non-zero.

Consider the case of an equilibrium system

$$m(\mathbf{r}, \mathbf{v}, t+1) = m(\mathbf{r}, \mathbf{v}, t). \tag{8}$$

Bound orbits on a lattice have finite period, since phase space is finite and maps onto itself. Thus the whole system is periodic with a large period which is the lowest common multiple of the periods of its constituent orbits. As time evolves, the particles representing $f$ move along the orbit, and so in a steady state the mass of each particle in a given orbit is a constant. Thus equilibrium solutions of (6) are specified by a list of closed orbits, each of which has a single value of $m$ associated with it. This is the lattice version of Jeans' theorem: the labels of the closed orbits correspond to the integrals of motion of the continuous case.

The resolution of the lattice is described by the velocity half-width $V$ and the radius $R$ (recall that the positions and velocities are integers on the lattice sites), as well as the number of resolution elements in the integerised accelerations **a**

$$A \sim \frac{V^2}{R}, \tag{9}$$

and the number of timesteps per orbit

$$T \sim \frac{R}{V}. \tag{10}$$

A high-resolution lattice model must have $\min(R, V, A, T) \gg 1$. If we scale $V \sim R^\zeta$ as we increase the resolution (i.e. as we increase $R$), then

$$A \sim R^{2\zeta - 1}, \qquad T \sim R^{1-\zeta}, \tag{11}$$

and we require that

$$\frac{1}{2} < \zeta < 1. \tag{12}$$

A reasonable compromise is therefore $\zeta \simeq 0.75$.

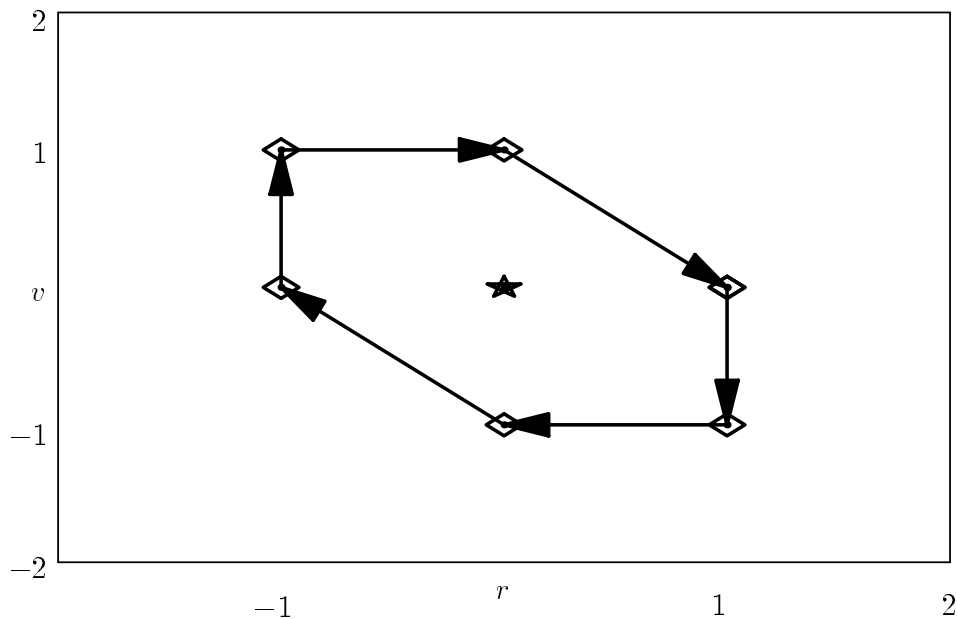

**Figure 1.** The two orbits in lattice phase for the example given in Section 3.1. One has period 6 (diamonds), and the other has period 1 (star).

# 3 Time-independent Systems

## 3.1 An Example of an Equilibrium

As a concrete example, consider a one-dimensional system with acceleration $a$ at position $r$ given by $a(1) = -1$, $a(0) = 0$, $a(-1) = 1$ and $a = 0$ elsewhere. There are two interesting closed orbits in this potential (see Figure 1): one with period $P = 6$, and a trivial one with $P = 1$. All particles not on these closed orbits are either unbound, and will leave the system, or are floating at $r = \text{const}$, $v = 0$. In equilibrium all the occupied sites on a given closed orbit have the same mass—$m_6$ for $P = 6$ and $m_1$ for $P = 1$, say—and the period of each orbit is equal to the number of lattice points on it.

We define the lattice density $\bar{m}(r)$ as

$$\bar{m}(r) = \sum_v m(r, v). \tag{13}$$

There are 2 orbits covering 3 values of $r$, and there are no non-trivial equilibria with $\bar{m}$ non-zero at any other radius. The set of possible equilibria is determined by a set of linear equations equivalent to solving the problem $\bar{m}(r) = m_0(r)$, where $m_0$ is a specified target density:

$$\begin{pmatrix} 2 & 0 \\ 2 & 1 \\ 2 & 0 \end{pmatrix} \begin{pmatrix} m_6 \\ m_1 \end{pmatrix} = \begin{pmatrix} m_0(-1) \\ m_0(0) \\ m_0(1) \end{pmatrix}. \tag{14}$$

The lattice density of an equilibrium is

$$(\bar{m}(-1), \bar{m}(0), \bar{m}(1)) = (2m_6, 2m_6 + m_1, 2m_6), \tag{15}$$

from which we see that all equilibria have the form

$$(\bar{m}(-1), \bar{m}(0), \bar{m}(1)) = (\rho, \rho + \delta, \rho) \tag{16}$$

where $\rho, \delta > 0$. There is no equilibrium with $\delta < 0$ (a bimodal mass distribution) or with $\bar{m}(-1) \neq \bar{m}(1)$ (an asymmetric mass distribution). The latter result has an analogue in continuous, one-dimensional stellar systems, which must be symmetric if the potential is symmetric.

The question of whether a given equilibrium is self-consistent, in the sense that it also satisfies (7), depends on the interparticle force law. If the interparticle force is too strong it will give $a = (2, 0, -2)$ near the centre. If the interparticle force is too weak, it will give $a = (0, 0, 0)$ near the centre.

## 3.2 An Algorithm for Finding Equilibria

In general, we would like to find a solution to (7) and (8) with $\bar{m}(\mathbf{r})$ equal to a specified density $m_0(\mathbf{r})$. We refer to $m_0$ as the target density, to the corresponding accelerations $\mathbf{a}_0$ as the target accelerations, and to the set of points for which $m_0 \neq 0$ as the target space. First we assign particles to the phase-space lattice, with $\bar{m} = m_0(\mathbf{r})$, and a specified velocity distribution (we normally use a Maxwellian with constant dispersion). We then carry out one of the two following procedures:

*3.2.1 Jeans Relaxation*
The name arises because at the end of the procedure the system is in an equilibrium, i.e. Jeans' theorem is satisfied; unfortunately the acceleration field is in general not self-consistent, i.e. Poisson's equation is not satisfied.
• We find the orbits of all the particles using the target accelerations, discarding those orbits that leave the target space. To do this it is not necessary to follow each orbit for its whole period. Rather it suffices to follow a particle until it either intersects another particle, or its starting point, or it leaves the target space.

• We try to find a solution of (8) using the surviving orbits. To do this we evolve the system using the target accelerations, and at each time step renormalise the masses of the particles so that $\bar{m}(\mathbf{r}) = m_0(\mathbf{r})$. The renormalisation is carried out by multiplying all the $m(\mathbf{r}, \mathbf{v})$ at each point on real space by a constant factor:

$$m(\mathbf{r}, \mathbf{v}) \to \frac{m_0(\mathbf{r})}{\bar{m}(\mathbf{r})} m(\mathbf{r}, \mathbf{v}). \qquad (17)$$

• The mass of all the particles on each orbit is then shared equally around the orbit. The lattice Jeans' theorem is always satisfied at this point, but not necessarily Poisson's equation. In this fashion mass is 'sifted' from orbit to orbit.

• The procedure is stopped if the integerised acceleration field, $\mathbf{a}$, of the current density, $\bar{m}(\mathbf{r})$, matches the target accelerations, $\mathbf{a}_0$, at points in real space where there are particles. We match the integerised acceleration field, but do not require that the density field matches the target density exactly.

The method described above is similar to those used by Schwarzschild and others in the orbit-based approach to finding equilibria in a continuous phase space. A fixed background potential is assumed, a set of particle orbits is determined and masses are assigned to the orbits in such a way as to recover the density $m_0$ that gives rise to the potential. Such problems may be re-cast as an underdetermined system of linear equations. Equation (17) is reminiscent of the Richardson-Lucy deconvolution algorithm (Richardson 1972, Lucy 1974).

The above procedure is generally found to converge well for very small systems (radius less than 3 or so), and for slightly larger systems that are almost spherical (radius about 4), but only for spherical systems if the radius is larger. Those cases that do not converge to a self-consistent equilibrium tend to converge to a steady state that is not self-consistent (i.e. does not have $\mathbf{a} = \mathbf{a}_0$) within a few tens of iterations. This number appears not to vary much with the spatial size of the target. For systems in two dimensions with radius around 4, there are less than a few hundred orbits, and the existence or otherwise of solutions to the corresponding set of linear equations can be demonstrated easily. It was found that all of the systems that failed to converge have no solution to the implied linear programming problem, but many systems which do converge have similarly ill-defined linear-programming counterparts. That there exist such systems is a consequence of the fact that many density fields can have the same integerised accelerations. It is possible that none of these density fields solves the implied linear programming problem in question, but that some of them are nonetheless in equilibrium.

*3.2.2 Poisson Relaxation*

If Jeans Relaxation does not converge in a few crossing times a supplementary procedure is adopted. We refer to it as Poisson relaxation because at the end of a timestep the acceleration field is exact and self-consistent, i.e. Poisson's equation is satisfied, but the system does not generally satisfy Jeans' theorem.

• The system is evolved with exact self-gravity (i.e. calculating the accelerations using equation 7). At each timestep $m$ is replaced with a weighted average across the timestep:

$$m(\mathbf{r}, \mathbf{v}, t+1) \leftarrow (1-\alpha)m(\mathbf{r}, \mathbf{v}, t) + \alpha m(\mathbf{r}, \mathbf{v}, t+1), \tag{18}$$

where $\alpha < 1$ is a positive constant. The averaging smears the mass around closed orbits, so that each closed orbit tends towards a state of equilibrium. The smoothing effect of (18) is diffusive with a characteristic timescale $\sim P^2/\alpha$, where $P$ is the period of the orbit. There is a steady decrease of the mass in orbits which are unbound, with characteristic timescale $1/\alpha$.

• Because the system is now self-gravitating, the set of closed orbits in general is different from that found in the Jeans relaxation stage. Some particles may in fact escape from the target region, and we have to decide what to do about them. Merely deleting those particles that escape would cause a loss of mass, and eventually the self-gravity of the whole system could become too weak to contain it. To remedy this problem, particles that leave the target region are deleted, and in addition $m$ is renormalised at each time step:

$$m(\mathbf{r}, \mathbf{v}) \to \left[1 + \epsilon \frac{(m_0(\mathbf{r}) - \bar{m}(\mathbf{r}))}{\bar{m}(\mathbf{r})}\right] m(\mathbf{r}, \mathbf{v}). \tag{19}$$

With $\epsilon = 1$ this is equivalent to (17); with $\epsilon < 1$ it exerts a weaker 'pressure' on the system, forcing it slowly towards $\bar{m} = m_0$, on a timescale $\sim 1/\epsilon$ when $\alpha = 0$. Any particle with a mass less than some small value, $f_{\min}$, is deleted. The factors of $\alpha$ in (18) and $\epsilon$ in (19) ensure that any particle not on a closed orbit rapidly loses its mass and is deleted. We parametrise $\alpha$ in terms of the half life of a particle $n_\alpha$, so that

$$\alpha = 1 - \left(\frac{1}{2}\right)^{1/n_\alpha}. \tag{20}$$

Experience shows that $n_\alpha = 16$, $\alpha = 0.042$, works well, and that if $\epsilon \gtrsim \alpha$ the convergence of the mass of a typical particle appears to be underdamped, so we use $\epsilon = 0.01$.

Equation (19) implies that the system is driven towards one with $\bar{m}(\mathbf{r}) = m_0(\mathbf{r})$. The timescale for this equilibrium to be reached is of order $P_{\max}^2/\alpha$, where $P_{\max}$ is the longest period in the system, which may be many times its crossing time. We do not wait for Poisson relaxation on its own to converge—the convergence may be slow and is towards a state which, in some cases, we know does not have an equilibrium. Instead, the Jeans relaxation is re-tried at regular intervals with the current value of $\bar{m}(\mathbf{r})$ as a target. There is no guarantee that the algorithm will terminate in an equilibrium, but many of the systems which fail to converge under Jeans relaxation alone in fact do converge when the two procedures are alternated (see Table 1). A cycle of the algorithm consists of a fixed number, $n_P$, of Poisson timesteps at which point the state of the system is saved. This is followed by another fixed number, $n_J$, of Jeans iterations using the density at the end of the Poisson steps as the value of $m_0$. At the beginning of a new cycle, the saved state is restored. We use $n_P = n_J = 60$.

*3.2.3 Results*
Most of the equilibria we have examined are based on the perfect ellipsoid (de Zeeuw 1985, Binney and Tremaine 1987) for which

$$m_0(\mathbf{r}) = \frac{C}{(1+s^2)^2},\tag{21}$$

where $C$ is a constant, and

$$s^2 = \sum_i \frac{r_i^2}{a_i^2},\tag{22}$$

with core radii $\{a_i\}$. The continuous density (21) extends all the way to infinity in each dimension, so we have to truncate it to fit on a finite mesh. This is done by setting $m_0$ equal to zero outside the ellipsoid given by $s^2 = \beta^2$, where $\beta$ is a constant. The target region thus has semi-axes $A_i = \beta a_i$. The initial distribution in velocity space fills a $d$-dimensional cube of side $V_0$ at each point in real-space, where $d$ is the number of spatial dimensions in the system. The initial masses are given by

$$m(\mathbf{r}, \mathbf{v}) = m_0(\mathbf{r}) \left(\pi d V_0^2\right)^{-d/2} \exp\left(-\frac{v^2}{dV_0^2}\right).\tag{23}$$

The constant $C$ is chosen in such a way that $|\boldsymbol{\nabla}\Phi| = 0.51$ is the smallest value of the acceleration around the edge of the region where $m_0 > 0$, so that $|\mathbf{a}_0| \geq 1$ throughout that region.

The interparticle force used to calculate $\mathbf{a}$ is Newtonian: between particles of mass $m_j$ and $m_k$ the force is

$$\mathbf{f}_{jk} = -m_j m_k \frac{(\mathbf{r}_j - \mathbf{r}_k)}{|\mathbf{r}_j - \mathbf{r}_k|^3}.\tag{24}$$

The results are summarised in Table 1. In each case, $\beta = 2.5$, and the real-space grid used was the smallest square or cube of side $2^n$ which would accomodate the system (with $n$ an integer). We will refer to the simulations by the semi-axes (column 1 in Table 1). The number of particles $N$ in the equilibrium is equal to the nuumber of non-empty lattice sites. The three-dimensional triaxial simulations $(6, 4, 7)$ and $(5, 4, 7)$ fail to converge (after 480 cycles). Normally a half cycle of Jeans relaxation is tried before the first full cycle because it tends to reduce the number of particles in the system. Simulations $(9, 10)$ through $(9, 12)$ are marked with an asterisk because this extra half cycle is omitted. Most of these simulations also converge if the extra half cycle is included, but simulation $(9, 10)$ is unusual because it fails to converge after 300 cycles. When the extra half cycle is omitted, it converges after only 2 cycles. Unfortunately, the same trick does not work for simulations $(6, 4, 7)$ and $(5, 4, 7)$.

Figure 2 shows the target density and the final equilibrium density for simulation $(4, 7)$. The two density fields do not match exactly because we only match the integerised accelerations. Figures 3 and 4 show the mass-weighted period distributions in some of the simulations. On the vertical axes are plotted the fractional mass found in the period bins along the horizontal axes.

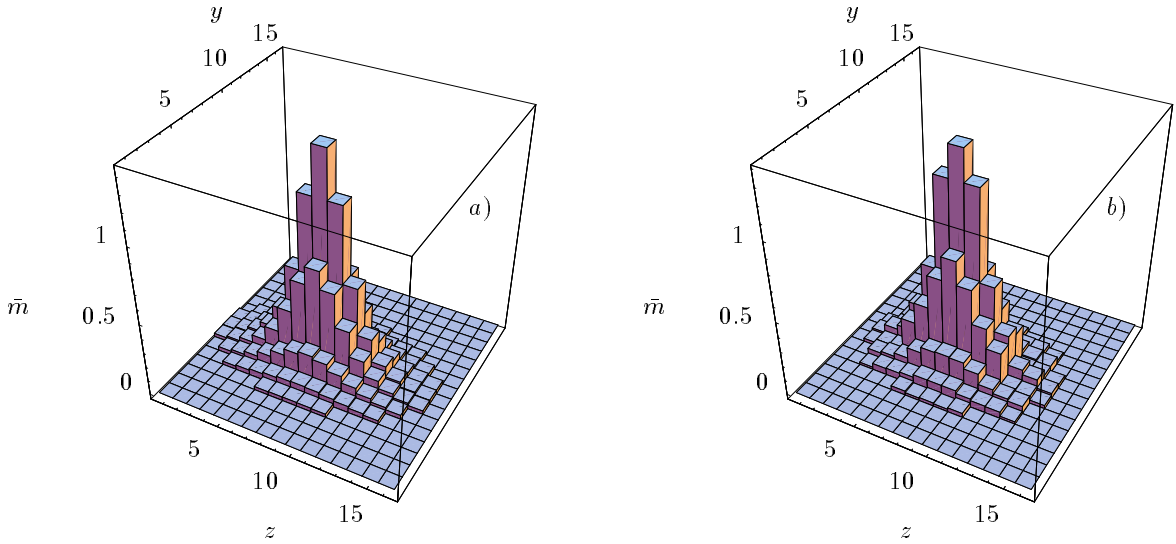

**Figure 2.** The target density (a), and the equilibrium density (b) found in simulation $(4, 7)$.

## 4 Linear Stability Analysis

Linear stability of lattice systems is not a useful concept: since the accelerations are integerized, they are unchanged if the perturbations are sufficiently small, so that orbits are unaffected by small perturbations. However, the equilibrium lattice system provides a novel tool for studying the linear stability of the continuous stellar system from which it is derived.

The analysis of linear stability of stellar systems is most easily tackled using action-angle variables (e.g. Kalnajs 1977, Toomre 1977). The lattice analogue of the action-angle system of variables is one in which each point in phase space, $(\mathbf{r}, \mathbf{v})$, is labelled by the orbit it belongs to, $j$, and the number of timesteps, $s$, required to reach that point from a given starting point on the orbit. The mass at a given site may be written $m = F_{js}$. The analog of Jeans' theorem (which states that the distribution function depends only on actions in an equilibrium system) is that the mass depends only on the orbit label, not the timestep, which may be stated recursively as

$$F_{js} = F_{j(s+1)}. \tag{25}$$

In contrast to continuous systems, every finite lattice phase space has action-angle variables because all orbits have finite periods. If the period of orbit $j$ is $P_j$, we may define the corresponding frequency,

$$\Omega_j = \frac{2\pi}{P_j}. \tag{26}$$

Now consider a perturbation to a continuous stationary stellar system, in which the distribution function $F \to F + f$, and the Hamiltonian $H \to H + h$. The linearised collisionless Boltzmann equation takes the form

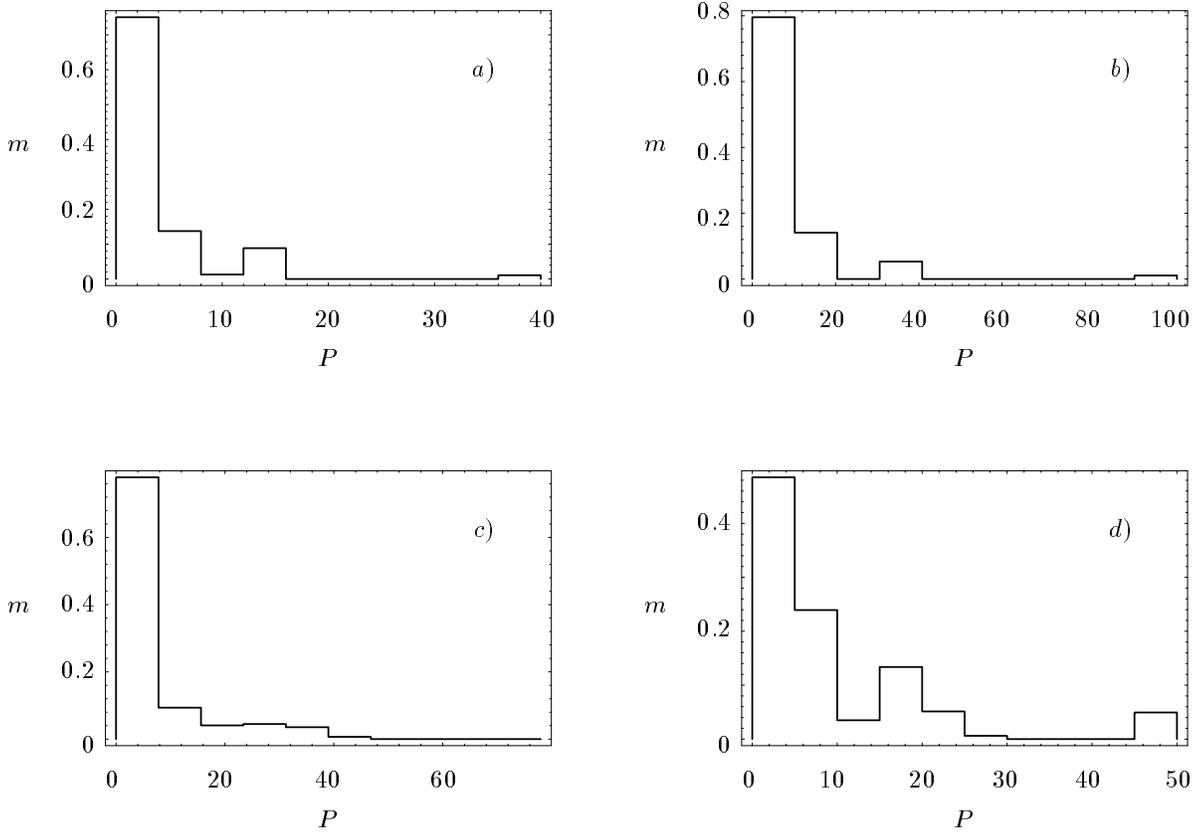

**Figure 3.** The mass-weighted period distributions in (a) simulation $(4,4)$; (b) simulation $(4,5)$; (c) $(4,6)$; and (d) $(4,7)$.

$$\frac{Df}{Dt} + \{h, F\} = 0, \qquad (27)$$

where $D$ denotes the Lagrangean derivative along the unperturbed orbit, and

$$\{h, F\} = -\mathbf{a} \cdot \frac{\partial F}{\partial \mathbf{v}} = -\mathbf{a} \cdot F_{\mathbf{v}} \qquad (28)$$

is the Poisson bracket, with

$$\mathbf{a} = -\boldsymbol{\nabla} h. \qquad (29)$$

The linearised Poisson equation is

$$\boldsymbol{\nabla}^2 h = \int f \, d\mathbf{v}. \qquad (30)$$

Equations (27) and (30) admit solutions with time dependence $\exp(-i\omega t)$.

We now approximate $f$ by its values at the lattice sites, $f_{js}(t)$. We can find an approximation to $F_{\mathbf{v}}$ from our lattice equilibrium by a simple centred difference formula:

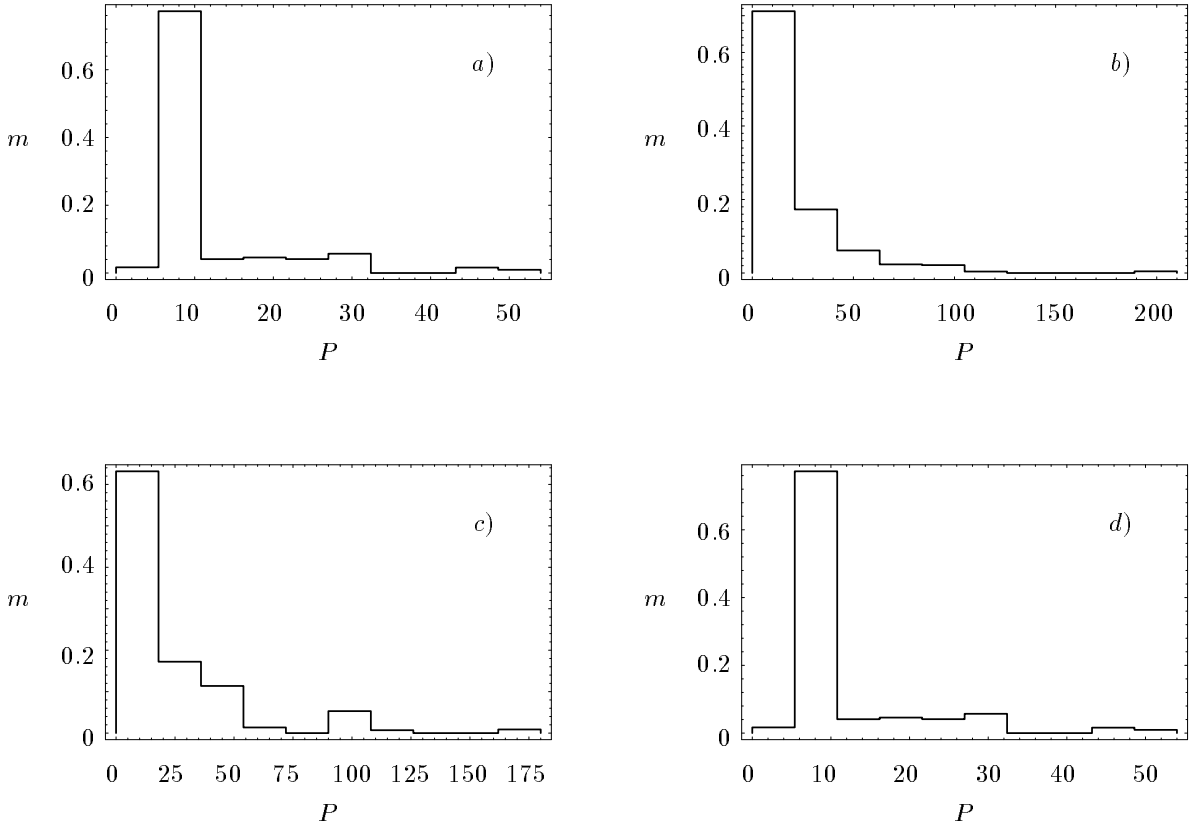

**Figure 4.** The mass-weighted period distributions in (a) simulation $(9,9)$; (b) $(9,12)$; (c) $(9,15)$; and (d) $(4,4,4)$

$$F_{\mathbf{v}}(\mathbf{r},\mathbf{v}) = \sum_{i=1}^{d} \hat{\mathbf{r}}_i \frac{\bar{m}(\mathbf{r},\mathbf{v}+\hat{\mathbf{v}}_i) - \bar{m}(\mathbf{r},\mathbf{v}-\hat{\mathbf{v}}_i)}{2}, \qquad (31)$$

where $\hat{\mathbf{r}}_i$ and $\hat{\mathbf{v}}_i$ are unit vectors in the $i$th co-ordinate direction. The acceleration $\mathbf{a}$ can be computed from $f_{js}$ in the same way that it was computed in Section 2, but without integerising. Then the lattice approximation to the evolution of a linear perturbation is given by

$$f_{j(s+1)}(t+1) = f_{js}(t) + \mathbf{a} \cdot F_{\mathbf{v}}. \qquad (32)$$

The solution to this equation yields the growth rate and the eigenfunction of the most unstable normal mode. In the remainder of this section we pursue the more difficult task of finding all the eigenfunctions and eigenfrequencies.

The linearised equations for the continuous system admit solutions with time-dependence $\exp(-i\omega t)$. We therefore expand $f_{js}$ in a Fourier series:

$$f_{js}(t) = \sum_{m=0}^{P_j-1} \hat{f}_{jm} \exp(im\Omega_j s - i\omega t). \qquad (33)$$

| $(A_1, A_2, A_3)$ | $V_0$ | $A_0$ | $N$ | $n_\alpha, \epsilon$ | $N_c$ |
|---|---|---|---|---|---|
| 3, 3 | 1 | 2 | 257 |  | 0 |
| 3, 4 | 1 | 2 | 297 | 1, .01 | 1 |
| 3, 5 | 1 | 2 | 389 | 16, .05 | 1 |
| 3, 6 | 1 | 2 | 661 | 16, .05 | 1 |
| 4, 4 | 1 | 2 | 501 |  | 0 |
| 4, 5 | 1 | 2 | 649 | 16, .1 | 1 |
| 4, 6 | 1 | 2 | 673 | 16, .01 | 1 |
| 4, 7 | 1 | 3 | 793 | 4, .01 | 14 |
| 9, 9 | 3 | 2 | 4529 |  | 0 |
| 9, 10 | 3 | 3 | 4507 | 4, .05 | 3* |
| 9, 11 | 3 | 2 | 4327 | 4, .05 | 27* |
| 9, 11 | 3 | 2 | 5223 | 16, .01 | 3* |
| 9, 12 | 3 | 2 | 5647 | 4, .05 | 2* |
| 9, 15 | 3 | 3 | 6707 | 4, .05 | 3 |
| 4, 4, 4 | 1 | 2 | 5163 |  | 0 |
| 6, 4, 6 | 1 | 2 | 12321 | 16, .01 | 1 |
| 5, 4, 7 | 1 | 2 | 0 | 16, .01 | > 480 |
| 6, 4, 7 | 1 | 2 | 0 | 16, .01 | > 480 |

**Table 1.** Perfect ellipsoidal equilibria. The first column gives the semi-axes of the region in which $m_0 > 0$. $N$ is the number of particles in the final equilibrium; an entry of zero represents a failure to find an equilibrium. $A_0$ is the largest component of $\mathbf{a}_0$. The parameters $n_\alpha$ and $\epsilon$ specify the parameters of the Poisson relaxation; a null entry indicates that Jeans relaxation converged on its own. The last column, $N_c$, is the number of Jeans-Poisson cycles required for convergence (* see text).

The lattice version of the Lagrangean derivative of $f$ is then rather simple:

$$\left.\frac{Df}{Dt}\right|_{js} = \sum_m i(m\Omega_j - \omega)\hat{f}_{jm} \exp(im\Omega_j s - i\omega t). \tag{34}$$

If we expand $\mathbf{a} \cdot F_{\mathbf{v}}$ in the same way, we can solve (27) immediately:

$$\hat{f}_{jm} = \frac{\widehat{(\mathbf{a} \cdot F_{\mathbf{v}})}_{jm}}{\omega - m\Omega_j}. \tag{35}$$

Now we turn to the the Poisson equation, and write it in the $(j, s)$ system as

$$\mathbf{a}_{kp} = \sum_{js} f_{js}\, \chi(\mathbf{r}_{kp} - \mathbf{r}_{js}), \tag{36}$$

where $\chi(\mathbf{r}_{kp} - \mathbf{r}_{js})$ is the interparticle force per unit mass. (Note that $\mathbf{a}$ is a function of $\mathbf{r}$ alone, but not of either $k$ or $p$ alone—as occurs when one has to solve Poisson's equation in action-angle co-ordinates.) Forming the dot product $\mathbf{a} \cdot F_{\mathbf{v}}$, we obtain after substituting from (33),

$$(\mathbf{a} \cdot F_{\mathbf{v}})_{kp} = \sum_{js}\sum_m \hat{f}_{jm}\, \exp(im\Omega_j s)(F_{\mathbf{v}})_{kp} \cdot \chi(\mathbf{r}_{kp} - \mathbf{r}_{js}) \tag{37}$$

(dropping the time dependence without loss of information). Approximating the Fourier dot product as a sum over $p$, we can pick out the $n$th component of $(\mathbf{a}\cdot F_\mathbf{v})_{kp}$ as follows:

$$(\widehat{\mathbf{a}\cdot F_\mathbf{v}})_{kn} = \frac{1}{P_k}\sum_p\sum_{js}\sum_m \hat{f}_{jm}(F_\mathbf{v})_{kp}\cdot\chi(\mathbf{r}_{kp}-\mathbf{r}_{js})\,\exp i\,(m\Omega_j s - n\Omega_k p). \quad (38)$$

Substituting (35) into (38) we obtain an eigenvalue equation for $(\widehat{\mathbf{a}\cdot F_\mathbf{v}})_{kn}$:

$$(\widehat{\mathbf{a}\cdot F_\mathbf{v}})_{kn} = \frac{1}{P_k}\sum_p\sum_{js}\sum_m \frac{(\widehat{\mathbf{a}\cdot F_\mathbf{v}})_{jm}}{\omega - m\Omega_j}(F_\mathbf{v})_{kp}\cdot\chi(\mathbf{r}_{kp}-\mathbf{r}_{js})\,\exp i\,(m\Omega_j s - n\Omega_k p). \quad (39)$$

We can re-write equation (39)

$$(\widehat{\mathbf{a}\cdot F_\mathbf{v}})_{kn} = \sum_{jm}\frac{(\widehat{\mathbf{a}\cdot F_\mathbf{v}})_{jm}}{\omega - m\Omega_j}A_{knjm}, \quad (40)$$

where

$$A_{knjm} = \frac{1}{P_k}\sum_{ps}(F_\mathbf{v})_{kp}\cdot\chi(\mathbf{r}_{kp}-\mathbf{r}_{js})\,\exp i\,(m\Omega_j s - n\Omega_k p) \quad (41)$$

can be evaluated using fast Fourier transforms.

Equation (40) is in the form

$$X_K = M_{KJ}X_J \quad (42)$$

with $K$ an integer assigned uniquely to each value of $(k,n)$. Hence equation (40) only has solutions when

$$|1 - M| = 0. \quad (43)$$

Thus we have reduced the problem of determining the eigenvalues of the linear stability problem to evaluating the roots of two polynomials (one for each of the real and imaginary parts of the Fourier transform 41) of degree

$$\sum_j P_j = N. \quad (44)$$

We now write

$$M_{KJ} = \Lambda^{-1}(\omega)_{JL}A_{KL} \quad (45)$$

where $\Lambda$ is the diagonal matrix $\omega - m\Omega_j$, and then equation (42) can be expressed as a linear eigenvalue problem:

$$[\Lambda(\omega)_{KJ} - A_{KJ}]Y_J = 0, \quad (46)$$

where $Y_J = \Lambda^{-1}(\omega)_{JL}X_L$. To avoid problems with resonances we can replace $m \to m + i\eta$ in $\Lambda(\omega)$, with $\eta$ real, and then take the limit as $\eta \to 0$. This is similar to the way the integral form of equation (42) is regularised in the analysis of continuous equilibria.

| $(P_j, m) \rightarrow$ $(P_k, n) \rightarrow$ | (6,0) | (6,1) | (6,2) | (6,3) | (6,4) | (6,5) | (1,0) |
|---|---|---|---|---|---|---|---|
| (6,0) | 0 | 0 | -0.00270633 | 0 | 0.00270633 | 0 | 0 |
| (6,1) | 0 | 0.289216 | 0 | -0.143796 | 0 | -0.144608 | 0 |
| (6,2) | 0 | 0 | 0.00378886 | 0 | 0.00189443 | 0 | 0 |
| (6,3) | 0 | 0 | 0 | 0 | 0 | 0 | 0 |
| (6,4) | 0 | 0 | -0.00189443 | 0 | -0.00378886 | 0 | 0 |
| (6,5) | 0 | 0.144608 | 0 | 0.143796 | 0 | -0.289216 | 0 |
| (1,0) | 0 | 0 | -0.00108253 | 0 | 0.00108253 | 0 | 0 |

**Table 2.** The imaginary part of the matrix $A$ for the simple equilibrium of Section 3.1.

As a simple example let us analyse the equilibrium from Section 3.1. When the interparticle potential is given by equation (24), an equilibrium exists with $m_6 = 0.0075$ and $m_1 = 0.5$. Table 2 lists the imaginary part of the matrix $A$ (the real part is independent and yields a nominally different set of eigen-frequencies). It is of block diagonal form: the top left $6 \times 6$ corner represents the self interaction of the $P = 6$ orbit, and the other non-zero contribution is the bottom row which represents the influence of the $P = 1$ orbit on each mode of the $P = 6$ orbit. (The transpose of the latter—the influence of the $P = 6$ orbit on the single mode of the $P = 1$ orbit—vanishes because $F_\mathbf{v} = 0$ at the centre.)

The diagonal elements of $\Lambda(0)$ are, in ascending order,

$$(0, 0, 1.0472, 2.0944, 3.14159, 4.18879, 5.23599) \qquad (47)$$

and the eigenfrequencies are, in ascending order,

$$\omega = (0, 0, 1.34222, 2.09819, 3.14159, 4.18500, 4.94097). \qquad (48)$$

The eigenfrequencies are all fairly close to, or exactly on, resonance. The introduction of $\eta$, which regularises equation (42), turns out not to affect the results, at least in the limit of small $\eta$. (This is also the case for the examples calculated below.) The closeness of the eigenfrequencies to resonance is not surpising in such a simple system, with only two short-period orbits—in larger equilibria, as the complexity of the system increases, more frequencies are off resonance. The neutral frequencies ($\omega = 0$) correspond to eigenmodes of uniform increase of the mass in each orbit. These can be seen to be neutral by symmetry in this very simple example. Note that there are no growing modes (frequencies with imaginary parts).

In Figure 5 we show the maximum growth rate $\gamma$ (the largest of the imaginary parts of all the eigenfrequencies) for the family of equilibria $(3, n)$ of the perfect ellipsoid analysed in Section 3.2.3. The period of a simple closed orbit along the minor axis is 6 in each case, which we may take as a measure of the dynamical time. All of these equilibria have some modes with imaginary frequencies, and are therefore formally unstable. The instability becomes more severe as the axis ratio increases, and the shortest growth time is a few dynamical times.

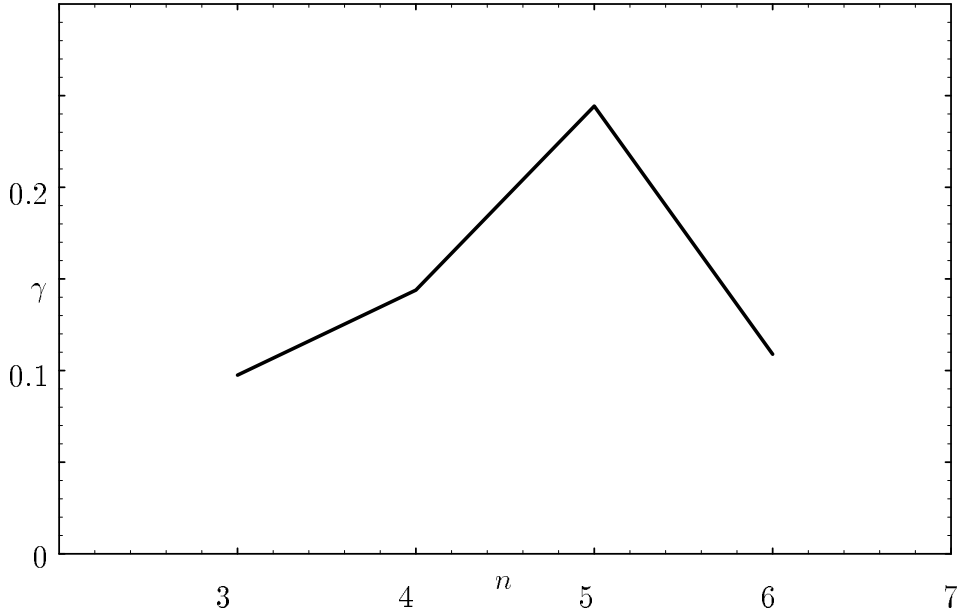

**Figure 5.** The maximum growth rate $\gamma$ of linear perturbations to the family of equilibria $(3, n)$.

# 5 Discussion

Does the lattice approach to stellar systems have advantages over continuous methods? The algorithm described here bears the closest relationship to the orbit-based methods of finding collisionless equilibria; however the storage required is proportional to the number of phase-space cells, rather than the number of position cells times the number of orbits. An advantage over orbit-based methods is that it also contains the elements of an initial-value approach, such as a continuous $N$-body code.

The lattice approach may be suitable for stability analysis of collisionless equilibria. The algorithm of Section 3 may be used to set up an exact equilibrium, and the analysis of Section 4 then applied. The eigenvalue problem is finite (although large); in cases where it is unmanageable, equation (32) could be solved numerically to give the fastest growing mode. Converting from Cartesian phase-space variables to action-angle variables is natural. There is no need to expand the system in terms of a set of co-ordinate space functions: the analysis is fundamentally based in phase space, in terms of direct orbit-orbit interactions.

Another area of potential interest is the study of equilibria in potentials which contain stochastic orbits in the continuous limit. The analogue of stochastic orbits in a lattice phase space is very long period orbits. The distribution of periods in an equilibrium may contain a clear signature of the transition to chaos, and thus might complement measures such as Liapunov exponents. In the present simulations there is an increase in the mass fraction of long period orbits as the deviation from axial symmetry becomes progressively more pronounced (Figures 3a-d, and Figures 4a-d). The perfect ellipsoids that they represent have only regular orbits in the continuum limit, but the high symmetry of the continuous systems is broken by the rectangular lattice.

The principal limitation of lattice stellar dynamics, like other phase-space methods, is the amount of storage required. In $d$ spatial dimensions the number of lattice sites is roughly

$$N \sim V^d R^d \sim R^{d(\zeta+1)}, \qquad (49)$$

where the exponent $\zeta$ is defined following equation (10). For a typical value $\zeta = 0.75$ (eq. 12),

$$N \sim R^{1.75 d}. \qquad (50)$$

In the simulations here the amount of computer memory required was not dominated by the storage of the particles, but for much larger simulations it would be. The dynamic range of the approach could be enhanced by using a non-rectangular lattice with spatially dependent resolution. Sparse equilibria, in which the majority of the lattice is empty, can be followed more easily, but in this case the lattice approach has no obvious advantage over standard $N$-body techniques.

# 6 Concluding remarks

Lattice stellar dynamics is most similar to phase-space methods for modeling solutions of the collisionless Boltzmann equation. The lattice approach has several advantages over traditional Eulerian methods of simulating the evolution of the phase-space fluid: the equations of motion are exactly symplectic, the effects of finite resolution are explicit, and the numerical method is easy to implement. For many applications we believe that the lattice approach is the phase-space method of choice. On the other hand, lattice stellar dynamics shares with other phase-space methods the limitation that storage quickly becomes unmanageable as the resolution is increased; thus its usefulness is probably restricted to systems with only a few phase-space dimensions (e.g. spherically symmetric or disc systems).

Lattice dynamics has the advantage over $N$-body experiments that it is possible to suppress relaxation completely and thus to construct exact steady-state systems.

Finally, the approach to the continuum limit in lattice stellar dynamics is different to that of other numerical methods, and hence may provide novel insights into the behaviour of real stellar systems.

We thank David Earn and Christophe Pichon for helpful discussions. This research was supported by NSERC and PPARC through a NATO fellowship. ST thanks the Killam Program and the Raymond and Beverly Sackler Foundation for support while this paper was being written.